\documentclass{amsart}
\usepackage{amssymb}
\usepackage{amsfonts}
\usepackage{epsfig}
\usepackage{graphicx}
\newcounter{eqnletter}[equation]
\catcode`\@=11

\newcommand{\half}{{\scriptstyle\frac{1}{2}}}
\newcommand{\ket}[1]{\big|#1\big\rangle}

\newcommand{\Tr}[1]{{\sf tr}\left(#1\right)}

\begin{document}
{\centerline {\LARGE {\bf Fock space methods and large $N$ }}} \vskip
1 cm \centerline {M.~Bonini, G.M.~Cicuta and E.~Onofri }
\vskip .5 cm {\small \centerline {{\it Dipartimento di Fisica, Univ. di Parma,
      v.~G.P.~Usberti 7A, 43100 Parma, Italy }} \centerline {{\it and
      INFN, Sezione di Milano Bicocca, Gruppo di Parma} \quad \quad } \centerline
{\small{\sf email}:\emph{ name(at)fis.unipr.it}}}
 
\vskip .7 cm {\centerline{\bf Abstract}}
Ideas and techniques (asymptotic decoupling of single-trace subspace, asymptotic operator algebras, duality and role of supersymmetry) relevant in current Fock space investigations of quantum field theories have very simple roles in a class of toy models.\\
\vskip .7 cm

Hamiltonian methods have a long history in the attempts to understand
the bound states spectrum of a strongly interacting relativistic
quantum field theory.  This is perhaps the hardest and most important
problem in a strongly interacting quantum field theory and analytic
and numerical efforts were devoted to invent reliable methods.  For
several years light-front quantization \cite{lc} was
a promising approach because of the very different nature of the ground state and some important simplifications of the operators occurring in the hamiltonian of non-abelian models.\\
The analysis of the large-$N$ limit, at t'Hooft coupling fixed, of the
models with $SU(N)$ or $U(N)$ invariance, provided additional
insights, indicating features of string theory in non-abelian gauge
models and the existence of symmetries, conserved quantum numbers and
operator algebras occurring only in the asymptotic theory, at
$N=\infty$, \cite{dalley}, \cite{alg}, \cite{lee}, \cite{hal}.  Much
work was devoted to the evaluation of the spectrum of the Hamiltonian
for states in the lowest representation of the group : the singlet
sector and the adjoint sector.

A color-singlet state of $n$ free bosons, with total momentum ${\vec
  P}$ is represented in the Fock space by a linear superposition of
states of the form
\begin{equation}
 \Tr{a^\dag({\boldsymbol k_1})\cdots a^\dag ({\boldsymbol k_{n_1}})} \; \Tr{ a^\dag({\boldsymbol p_1})\cdots a^\dag ({\boldsymbol p_{n_2}})} \cdots  \Tr{ a^\dag({\boldsymbol q_1})\cdots a^\dag ({\boldsymbol q_{n_s}})} \,\ket{0}
 \label{t.1}
\end{equation}
where $\Tr{}$ denotes the trace on $U(N)$ colour indices,
$n=n_1+n_2+\cdots +n_s$, ${\boldsymbol P}= \sum {\boldsymbol k_j}+\sum
{\boldsymbol p_j}+\dots \sum {\boldsymbol q_j}$ and $a^\dag
({\boldsymbol k})$ are creation operators. These states are called
multi-trace states.  Matrix-valued operators may be written in terms
of the group generators $ a({\boldsymbol k})=\sum_a \lambda_a
a_a({\boldsymbol k})$, $ a^\dag({\boldsymbol k})=\sum_a \lambda_a
a_a^\dag({\boldsymbol k})$, and all matrix-related coefficients are
efficiently evaluated by graphic methods \cite{cvi, mt}. Several
remarkable properties were found in the large-$N$ limit.  Operators
nor\-mal\-ly--or\-der\-ed inside a single-trace, that is of the form
$\gamma=N^{-(r+s-2)/2}\;\Tr{a^\dag({\boldsymbol k_1})\cdots a^\dag
  ({\boldsymbol k_r})\,a({\boldsymbol p_1})\cdots a({\boldsymbol p_s})}$, 
acting on single-trace states generate single-trace
states \cite{lee}, provided $r \geq 1$ and $s \geq 1$. Then if the
Hamiltonian is a linear combination of such operators, the subspace of
the Fock space spanned by single-trace states is invariant under the
action of the Hamiltonian. These operators and analogous ones
involving fermion operators act on single-trace states in a way
reminiscent of the coupling of strings. Single-trace states like
$\ket{n} = N^{-n/2}n^{-1/2} \Tr{ a^\dag({\boldsymbol k_1})\cdots
  a^\dag ({\boldsymbol k_{n}})}\,\ket{0}$ form an orthonormal basis in
the color-singlet and single-trace Fock space \cite{thorn}.
Multi-trace states like in eq.(\ref{t.1}) provide an orthonormal basis
in the color-singlet Fock space at $N=\infty$ \cite{thorn}.

Recently G.~Veneziano and J.~Wosiek \cite{past} suggested a
supersymmetric model in $D=1$ space-time dimension, that is a matrix
quantum mechanical model.  It is not surprising that the model is
analytically solvable in the large-$N$ limit in several sectors of the
Fock space and reliable numerical evaluations may be performed in
other sectors. Several features of the bound states are very
interesting.

The goal of this letter is to use results derived in recent analysis
of the bosonic sector of the model \cite{rob} and in a simple
generalization presented here, to comment on the properties of V-W
hamiltonian with a view to the role they may have in models in
more realistic space-time dimension.

\vskip .25 cm
In the bosonic sector the Hamiltonian of the V-W model is
\begin{equation}
H=\Tr{ a^\dag a+g\left(a^{\dag 2}a+a^\dag a^2\right)+g^2a^{\dag 2}a^2},\quad  \lambda=g^2N 
\label{h.1}
\end{equation}
In the single trace sector of the singlet states, for large $N$, the Hamiltonian is a tridiagonal real symmetric infinite matrix.
The bound states spectrum was analytically and numerically solved in the large-$N$ limit for every $\lambda\geq 0$ and it presents remarkable features:\\
\begin{itemize}
\item the infinitely many eigenvalues of the discrete spectrum of the model, for $0 \leq \lambda<1$ decrease in a monotonous way as $\lambda$ increases and all vanish at $\lambda=1$, where a phase transition occurs. For $\lambda >1$ one eigenvalue remains at zero energy, it is a new ground state, and the infinitely many eigenvalues 
increase in a monotonous way as $\lambda$ increases.
\item The non-zero eigenvalues at the pairs of values $\lambda$ and $1/\lambda$ are related by a duality property 
\begin{equation}
\frac{1}{\sqrt{\lambda}}\left(E_n(\lambda)-\lambda)\right)=
\sqrt{\lambda}\left(E_n\left(\frac{1}{\sqrt{\lambda}}\right)-\frac{1}{\lambda}\right)
 \label{h.2}
\end{equation}
\end{itemize}
Let us consider the trivial generalization of eq.(\ref{h.1}) by
allowing two coupling constants
\begin{equation*}
H=\Tr{ a^\dag a + g_3\,(a^{\dag 2}a+a^\dag a^2) + g_4^2\,a^{\dag 2}a^2}\,,\; \sqrt{\lambda_3} = g_3 \sqrt{N}\,,\; \lambda_4=g_4^2 N
\qquad \qquad
 \label{h.3}
\end{equation*}

Here too the single-trace sector in the Fock space decouples in the
large-$N $ limit and the hamiltonian is represented by the tridiagonal
real symmetric infinite matrix $H(\lambda_3,\lambda_4)$
\begin{eqnarray}
&&H_{j,j+1}=H_{j+1,j}=\sqrt{\lambda_3}\sqrt{j(j+1)}\,, \nonumber\\
&&H_{j,j}=\left(1+\lambda_4(1-\delta_{1j})\right)j \,, \;\; 
 j=1, 2,\ldots
 \qquad \qquad
 \label{h.3b}
\end{eqnarray}
 The eigenvalue equation $H {\boldsymbol x}=E {\boldsymbol x}$ is a system of recurrent relations which translates into an easy differential 
equation for the generating function $G(z)=\sum_1^\infty y_j z^j$ where $x_j=y_j\sqrt{j}$.
\begin{eqnarray*}
&&\sqrt{\lambda_3}\,\omega(z)G'(z)-E\,G(z)-\left(z\lambda_4+\sqrt{\lambda_3}\right)G'(0)=0 \,,  \nonumber\\
&&{\rm where}\quad \omega(z)=z^2+1+z(1+\lambda_4)/\sqrt{\lambda_3}
 \label{h.5}
\end{eqnarray*}
If $(1+\lambda_4)^2-4\lambda_3>0$, $\omega(z)$ has $2$ distinct real
roots. The closest to the origin $z=0$ translates into the asymptotic
behaviour of the coefficients $x_j$. The integration constant of the
differential equation may be chosen to kill the closest singularity
then obtaining an exponentially decreasing sequence $x_j$, 
hence normalizable bound states.

The region of the quadrant $\lambda_3 \geq 0$ and $\lambda_4 \geq -1$ where the spectrum is discrete lies above the parabola of equation 
$4\lambda_3=(1+\lambda_4)^2$, shown by the black line in the plot.

 \epsfig{file=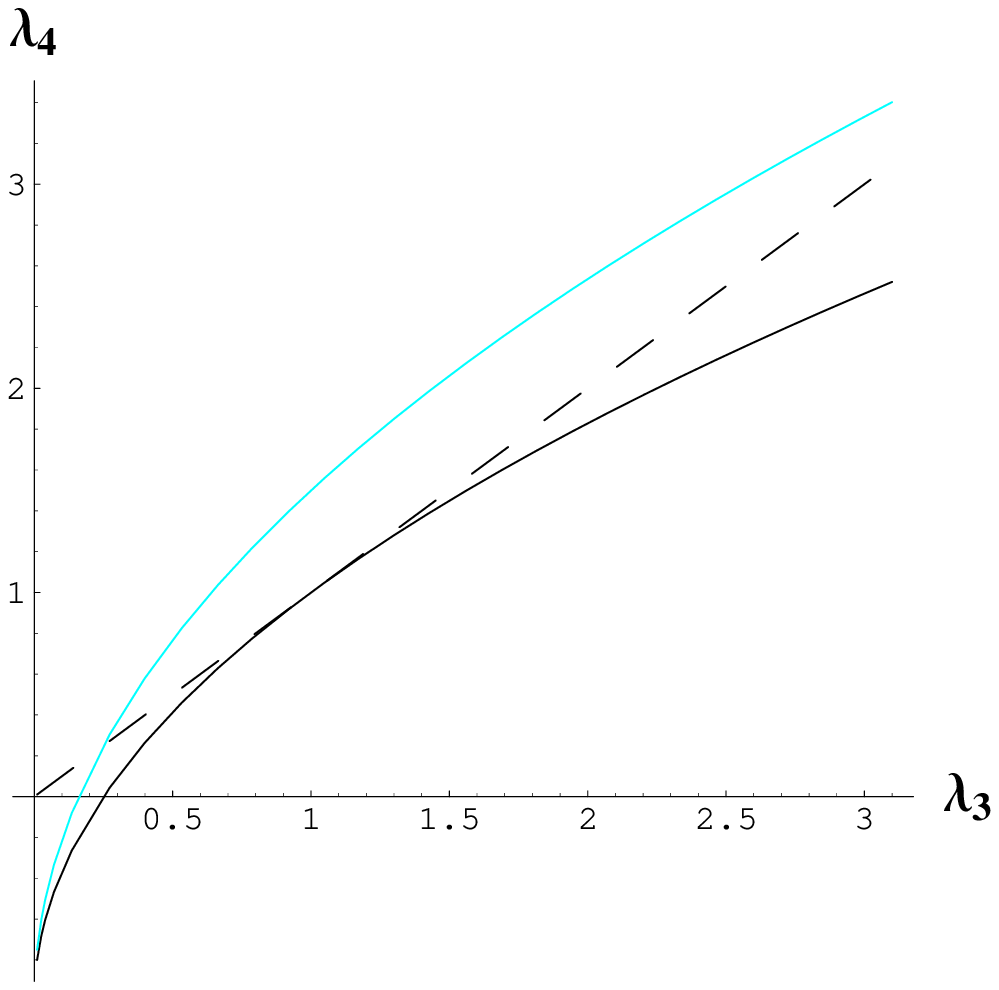,height=6cm}\\
 Fig.1. The line $\lambda_3=\lambda_4$ corresponds to the V-W model.\\

The solution has a compact form in terms of the variables $\sigma$ and $\eta$
  \begin{equation*}
    \sigma=\frac{ 1+\lambda_4- \sqrt{(1+\lambda_4)^2-4\lambda_3}}{1+\lambda_4+\sqrt{(1+\lambda_4)^2-4\lambda_3}} 
    \quad, \quad \eta=\frac{\lambda_4(1+\lambda_4)+\lambda_4 \sqrt{(1+\lambda_4)^2-4\lambda_3}-2\lambda_3}{2\lambda_4 \sqrt{(1+\lambda_4)^2-4\lambda_3}}
    \qquad \qquad
\end{equation*}
The eigenvalues $E_n$ of normalizable states are the infinitely many solution of the equation
\begin{eqnarray*}
  && F\left( \alpha,1;1+\alpha; \sigma \right)=\eta \;, \quad 0<\sigma <1
  \;, \qquad {\rm where} \nonumber\\
  &&F\left( \alpha,1;1+\alpha; \sigma \right)=1+\alpha \sum_{k=1}^\infty \frac{\sigma^k}{\alpha + k}\;, \quad \alpha=-\frac{E}{\sqrt{(1+\lambda_4)^2-4\lambda_3}}
\end{eqnarray*}
It is easy to check that the hypergeometric function $F\left(
  \alpha,1;1+\alpha; \sigma \right)$ satisfies the translation
identity
 \begin{equation}
   F\left( \alpha,1;1+\alpha; \sigma \right)=1+\frac{\alpha \,\sigma}{1+\alpha}F\left( \alpha+1,1;2+\alpha; \sigma \right)
  \label{h.40}
\end{equation}

The plot shows the Hypergeometric function $F\left( \alpha,1;1+\alpha;
  \sigma \right)$ versus $\alpha$ for fixed $\sigma$ and
$-6 < \alpha <2$.\\

 \epsfig{file=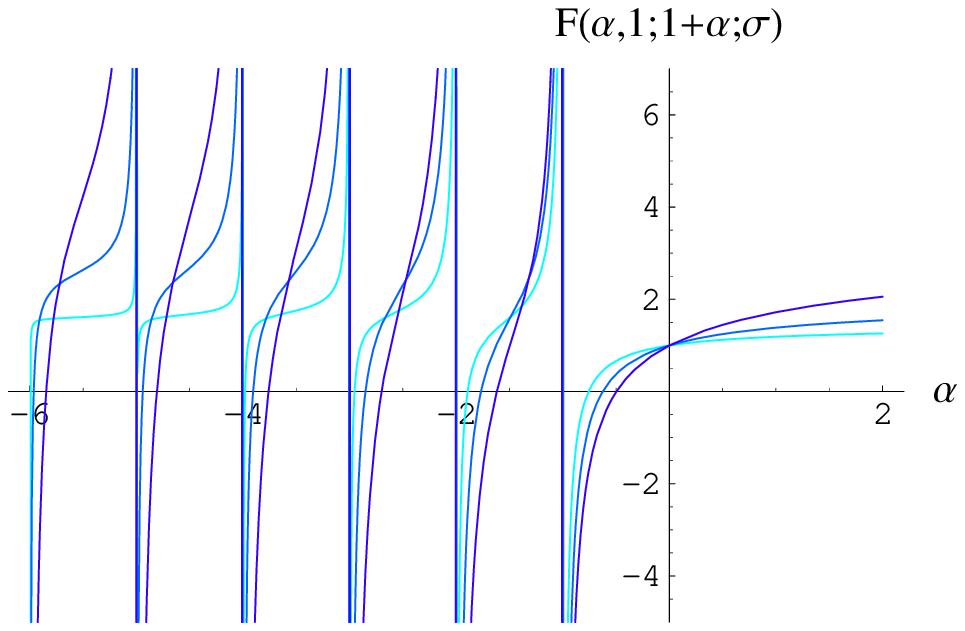,height=6cm}\\
 Fig.2. 
The graph shows $F\left( \alpha,1;1+\alpha; \sigma \right)$ for $3$ values $\sigma=0.3,\,0.5,\,0.7$ depicted with increasingly dark color.\\

In the space of parameters corresponding to bound states, let us
consider the parabolas of equation $\lambda_3= {\bar \sigma} \left(
  \frac{1+\lambda_4}{1+{\bar \sigma}}\right)^2$, see Fig.1. At each point of the
space of coupling constants corresponding to the discrete spectrum
there is a unique ${\bar \sigma}$. For a given parabola, it is easy to
describe the spectrum as we move along it from $\lambda_4=0$ to
$\lambda_4=\infty$. Indeed along this path, the value of $\eta$ increases in a monotonous way from $\eta=-\infty$ to $\eta=0$ at the point where it first crosses the line $\lambda_4=\lambda_3$, next it  reaches $\eta=1$ at the second crossing with the same line, and continues increasing up to its asymptotic limit $\eta=1/(1- {\bar \sigma}^2)$.\\
At $\eta=-\infty$ the eigenvalue equation has infinitely many solution
of the form $\alpha_n=-n+\epsilon_n$ where $n=1,2,\cdots$ and
$\epsilon_n$ are small positive numbers. That is $E_n=n(1-{\bar
  \sigma})/(1+{\bar \sigma})-\epsilon_n$. As $\eta$ increases all
roots $\alpha_n$ move in a monotonous way to the right, that is each
$E_n$ decrease. At $\eta=1$, the value of $\alpha_1=0$ and
$E_1=0$. This bound state becames degenerate with the vacuum state
$|0>$ of the Fock space, which has $E_0=0$. As $\eta$ increases beyond
$\eta=1$,
$\alpha_1$ increases to positive values and $E_1$ has {\bf increasingly negative} values \cite{some}.\\
At the first crossing $\lambda_3=\lambda_4=\lambda<1$, $\eta=0$,
${\bar \sigma}=\lambda$ and we may consider the roots $\alpha_n+1$ of
the eigenvalue equation $F(\alpha_n+1,1 ;\,\alpha_n+2
;\,\lambda)=0$. Because of the translation identity eq.(\ref{h.40})
they are simply related to the roots $\alpha_n$ of the eigenvalue
equation at the second crossing $\lambda_3=\lambda_4=\lambda>1$,
$\eta=1$, ${\bar \sigma}=1/\lambda$, which is
$F(\alpha_n,1 ;\,\alpha_n+1 ;\,1/\lambda)=1$. This is the V-W duality of eq.(\ref{h.2}).\\

This picture looks very different from the V-W spectrum but, of
course, it is fully compatible : in the V-W model the unique coupling
moves along the line $\lambda_4=\lambda_3$ and it touches at $\lambda=1$ the boundary of normalizable eigenstates.\\

We now briefly describe the solution of the $2$-couplings model by use
of a non--compact Lie algebra which arises in the large $N$ limit. 

Let us consider the Hamiltonian ${\tilde H}(\alpha, \beta)$
  \begin{eqnarray*}
    &&{\tilde H}(\alpha, \beta)=\alpha \, D + \half \beta (J_+ +J_-) \;, \qquad \nonumber\\
    && D_{ij}=j \delta_{ij} \;, \;\; (J_+)_{ij}=\sqrt{j(j+1)}\delta_{i,j+1} \;, \qquad J_-=(J_+)^\dag \qquad \nonumber\\
    &&J_\pm=J_x\pm iJ_y \;, \;\;
    \alpha=1+\lambda_4 \;, \;\; \beta=2\sqrt{\lambda_3}
    \qquad \qquad
  \label{h.80}
\end{eqnarray*}

The Hamiltonian ${\tilde H}(\alpha, \beta)$ differs from the
asymptotic Hamiltonian $H(\lambda_3, \lambda_4)$ given in
eq.(\ref{h.3b}) only for one matrix element ${\tilde
  H}_{11}=H_{11}+\lambda_4$. One easily computes the commutators
$[D,J_{\pm}]=\pm J_{\pm}$ and $[J_+,J_-]=-2D$, showing that
$\{D,J_+,J_-\}$ form a basis for the Lie algebra $SO(2,1)$ in the
degenerate Bargmann's discrete series representation $\mathcal
D_+^{n}$, ($n=0$),characterized by a vanishing Casimir operator. Hence
the generator $J_x$ has a continuous spectrum filling the whole real
axis \cite{Bargmann1947}.

Previous results about the spectrum can be re--derived as follows: if
$\alpha>\beta$, one writes
  \begin{equation*}
    \frac{1}{ \sqrt{\alpha^2-\beta^2} } {\tilde H}=\cosh y \,D+\sinh y \,J_x  \;, \;\; \cosh y=\frac{\alpha}{\sqrt{\alpha^2-\beta^2}}
  \label{h.81}
\end{equation*}
and a unitary operator $U$ (a boost) exists such that $U
\frac{1}{\sqrt{\alpha^2-\beta^2} } {\tilde H}\,U^{-1}=D$ and the
spectrum of ${\tilde H}$ is simply $E_n=n\sqrt{\alpha^2-\beta^2}$.

If $\alpha<\beta$, upon writing
 \begin{equation*}
   \frac{1}{ \sqrt{ \beta^2 -\alpha^2} } {\tilde H}=\sinh y \,D+\cosh y \,J_x \;, \;\; \cosh y=\frac{\beta}{\sqrt{\beta^2 -\alpha^2 }}
  \label{h.82}
\end{equation*}
a unitary operator boosting ${\tilde H}$ to $J_x$ exists, hence
${\tilde H}$ has a continuous spectrum.  On the border $\alpha=\beta$,
${\tilde H}$ coincides with a light--cone generator which has a
continuous (positive) spectrum \cite{Bargmann1947}.

Finally the spectrum of the asymptotic Hamiltonian $H(\lambda_3,
\lambda_4)$ may be computed from the spectrum of ${\tilde H}$ by the
method outlined in the appendix of \cite{rob}, based on an exact
perturbation formula (rank--one perturbation).

Notice that the operators closing the $SO(2,1)$ algebra represent the
restriction to the single--trace states of more general operators
acting on general singlet states. For instance $J_+\to \Tr{a^{\dag
    2}a}/\sqrt{N}$. The operators $(H,J_+,J_-)$ close a Lie algebra up
to terms of order $1/N$, hence the spectra can be discussed as arising
from a dynamical symmetry breaking.

 Let us summarize our conclusions :
 \begin{itemize}
 \item Hamiltonian models where each operator has the form ${\rm
     Tr}\,[(a^\dag)^n \,a^m]$ with $n \geq 1$ and $m \geq 1$ leave
   each sector of k-trace states invariant in the large-$N$ limit.  By
   representing the hamiltonian in the basis of single-trace states,
   one obtains a real symmetric fixed-width band matrix. The
   eigenvalue equation is a system of recurrent equations which
   usually allows analytic solution.  Still multi-trace singlet
   sectors are not irrelevant because the mass of bound states in
   these sectors is similar to that of the single-trace states.  As
   indicated in \cite{rob} the V-W bosonic hamiltonian may be
   evaluated in each sector obtaining the same eigenvalues, therefore
   changing (in infinite way) the degeneracy of eigenvalues found in
   the single-trace analysis.
 \item In the simple models in $D=1$ supersymmetry is not necessary
   for the asymptotic decoupling of the single-trace sector nor for
   the exact asymptotic solution of the model. However the duality
   property of the spectrum has a striking simple form only on the
   susy line $\lambda_3=\lambda_4$.
 \item Light-front quantization seems relevant for realistic models,
   that is in space-time dimension $1<D \leq 4$, because it makes
   possible to represent a local Hamiltonian in a partial normally
   ordered form \cite{hal} such that the single-trace sector of Fock
   space may be asymptotically invariant. This is a practical
   necessity for the Tamm-Dancoff approach. Exact or approximate
   algebras of the operators which occur in the asymptotic Hamiltonian
   provide a precious tool for the understanding of the spectrum.
 \item When considering a local quantum mechanical hamiltonian $H={\rm
     Tr}[p^2+x^2+V(x)]$ in the large-$N$ limit, the Fock space methods
   seems inappropriate: one cannot avoid operator terms that couple
   the single-trace states to multiple-trace states in leading
   order. This makes any evaluation restricted to the single-trace
   Fock states totally unreliable \cite{rob}\cite{bre}.
 \item At large $N$ a new dynamical symmetry shows up, simplifying the
   calculation of the spectrum. The interplay of this symmetry with
   supersymmetry as in V--W model deserves further study.
\end{itemize}

We add a last comment to indicate that the bosonic two-couplings model discussed in this letter is the asymptotic generic form of infinitely many Hamiltonians.\\
Let us consider the operators $A_j$, $A_j^\dag$, $D_j$, $j=1,2,\cdots$
  \begin{eqnarray*}
&&A_j^\dag =\frac{1}{N^{j-1/2}}\,\Tr{a^\dag (a^\dag a)^j}
\;, \;\;
A_j =\frac{1}{N^{j-1/2}}\,\Tr{ (a^\dag a)^j a}
\;, \nonumber\\
&&D_j=\frac{1}{N^{j-1}} \,\Tr{(a^\dag a)^j}
\qquad \qquad
  \label{h.60*}
\end{eqnarray*}
It is easy to verify that asymptotically they leave the sector of the
single-trace states invariant
\begin{eqnarray*}
&&A_j \,\ket{n} \sim \sqrt{n(n-1)}\,\ket{n-1} +O(1/N)\;, \nonumber\\
&&A_j^\dag \,\ket{n} \sim \sqrt{n(n+1)}\,\ket{n+1} + O(1/N)\;, \nonumber\\
&&D_j\,\ket{n} \sim n\,\ket{n} + O(1/N)\qquad 
\qquad \qquad
  \label{h.61}
\end{eqnarray*}
Then all the Hamiltonians
\begin{equation*}
  H= \Tr{ a^\dag a} + \sum_{j\ge1} g_{j,3}\;\Tr{A_j^\dag +A_j} + \sum_{j\ge2} g_{j,4}\;\Tr{D_j}  \label{h.62}
\end{equation*}
asymptotically  leave  the sector of the single-trace states  invariant and in this sector are all represented by the
 tridiagonal real symmetric infinite matrix  $H(\lambda_3,\lambda_4)$
in eq.(\ref{h.3b}) with $\lambda_3=\sqrt{N}\sum_j g_{j,3}$ and
$\lambda_4=N\sum_j g_{j,4}$.
The sums may be finite or infinite.

\begin{center}
{\bf Acknowledgments}
\end{center}
E.O. would like to thank G.~Veneziano, J.~Wosiek and M.~Trzetrzelewski
for interesting discussions.


\begin{thebibliography}{99}
\bibitem{some} Some readers may appreciate understanding the model without need of extensive numerical work.
\end{thebibliography}
\end{document}